\documentclass{pasj01}

\usepackage{color}
\usepackage{url}
\usepackage{ulem}
\usepackage{multirow}
\usepackage{longtable}

\usepackage{graphicx}
\usepackage{epstopdf}

 \usepackage{natbib}

\begin{document} 
\Received{}
\Accepted{}

\title{The sign of active galactic nucleus quenching in a merger remnant with radio jets}

\author{Kohei \textsc{Ichikawa}\altaffilmark{1, *}}%
\altaffiltext{1}{National Astronomical Observatory of Japan, 2-21-1 Osawa, Mitaka, Tokyo 181-8588, Japan}
\email{kohei.ichikawa@nao.ac.jp}

\author{Junko \textsc{Ueda}\altaffilmark{1}}

\author{Megumi \textsc{Shidatsu}\altaffilmark{2}}
\altaffiltext{2}{MAXI team, RIKEN, 2-1 Hirosawa, Wako, Saitama 351-0198, Japan}

\author{Taiki \textsc{Kawamuro}\altaffilmark{3},}
\altaffiltext{3}{Department of Astronomy, Kyoto University, Kitashirakawa-Oiwake-cho, Sakyo-ku, Kyoto 606-8502, Japan}

\author{Kenta \textsc{Matsuoka}\altaffilmark{3}}

\KeyWords{galaxies: active --- galaxies: nuclei --- infrared: galaxies}

\maketitle

\begin{abstract}
We investigate optical, infrared, and radio active galactic nucleus (AGN) signs in the merger remnant Arp~187, 
which hosts luminous jets launched in the order of $10^5$~yr
ago but whose present-day AGN activity is still unknown.
We find AGN signs from the optical BPT diagram and infrared \textsc{[Oiv]}25.89~$\mu$m line, 
originating from the narrow line regions of AGN.
On the other hand, \textit{Spitzer}/IRS show the host galaxy dominated spectra, 
suggesting that the thermal emission from the AGN torus is considerably small or already diminished.
Combining the black hole mass, the upper limit of radio luminosity of the core, and the fundamental plane of the black hole
 enable us to estimate X-ray luminosity, which gives $<10^{40}$~erg~s$^{-1}$.
Those results suggest that the AGN activity of Arp 187 has already been quenched, but the narrow line region is still alive owing to the time delay of emission from the
past AGN activity.
\end{abstract}

\section{Introduction}

A link between active galactic nuclei (AGN) and major galaxy mergers 
has been seen in luminous infrared galaxies \citep[e.g.,][]{san88}.  
Radial streaming motions during a merging event 
can efficiently feed gas to the central black hole (BH), 
and trigger AGN in late-stage mergers \citep[e.g.,][]{hop06, nar08}.  
The final stages of mergers are accompanied by a blow-out phase, 
expelling gas and dust into the intergalactic medium 
and quenching the star formation \citep[e.g.,][]{hop08}.  
Recent numerical simulations that incude merger-driven fueling and ``AGN feedback'' 
reproduce several observed properties of the galaxy populations, 
such as the BH mass--bulge mass correlation 
or the rapid build--up of massive spheroids \citep[e.g.,][]{spr05, dim05}.  
Therefore, AGN feedback has gradually been recognized 
as an important process in the evolution of galaxies.

\begin{figure*}
\begin{center}
\includegraphics[width=7.0cm]{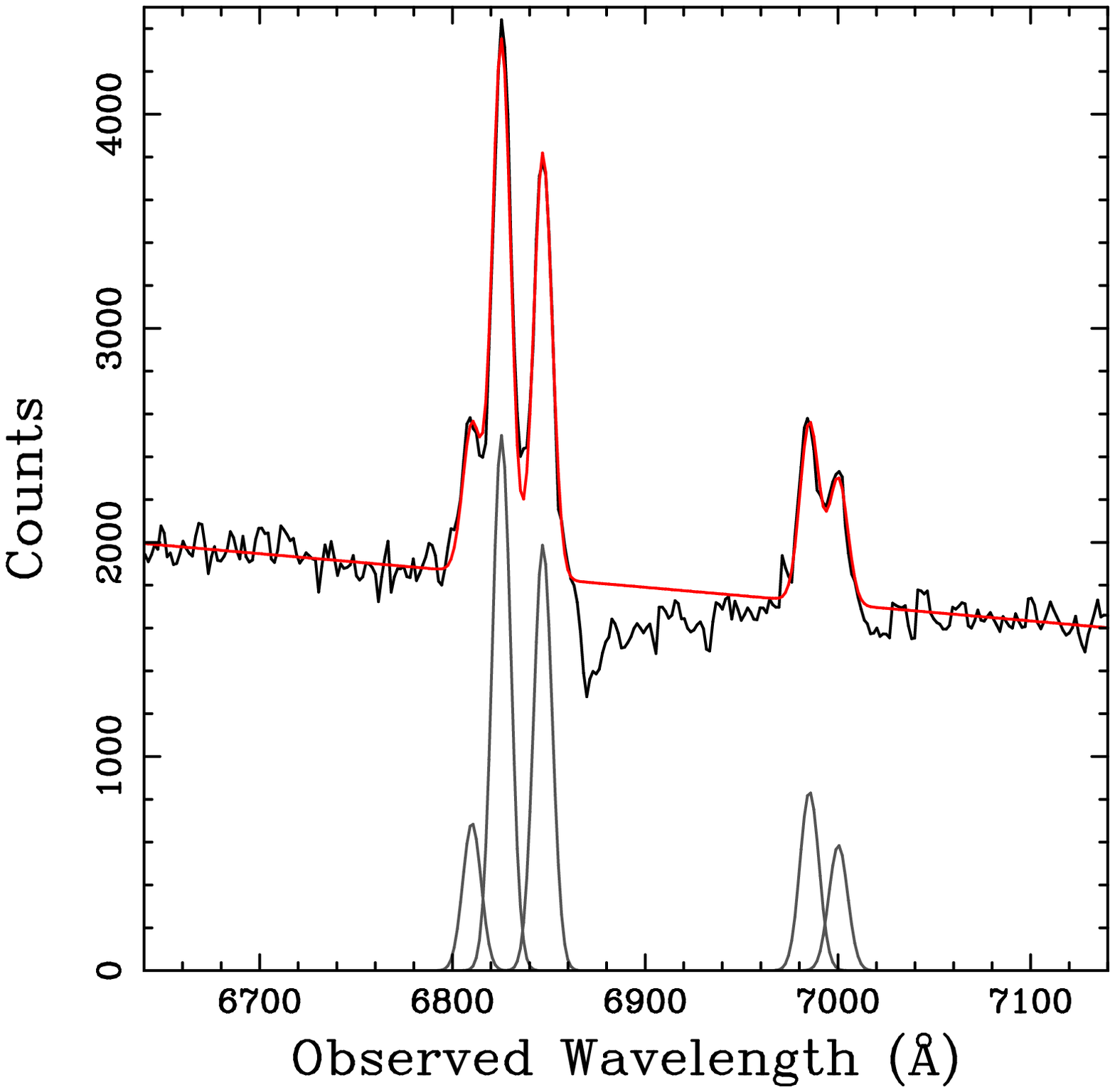}
\includegraphics[width=7.0cm]{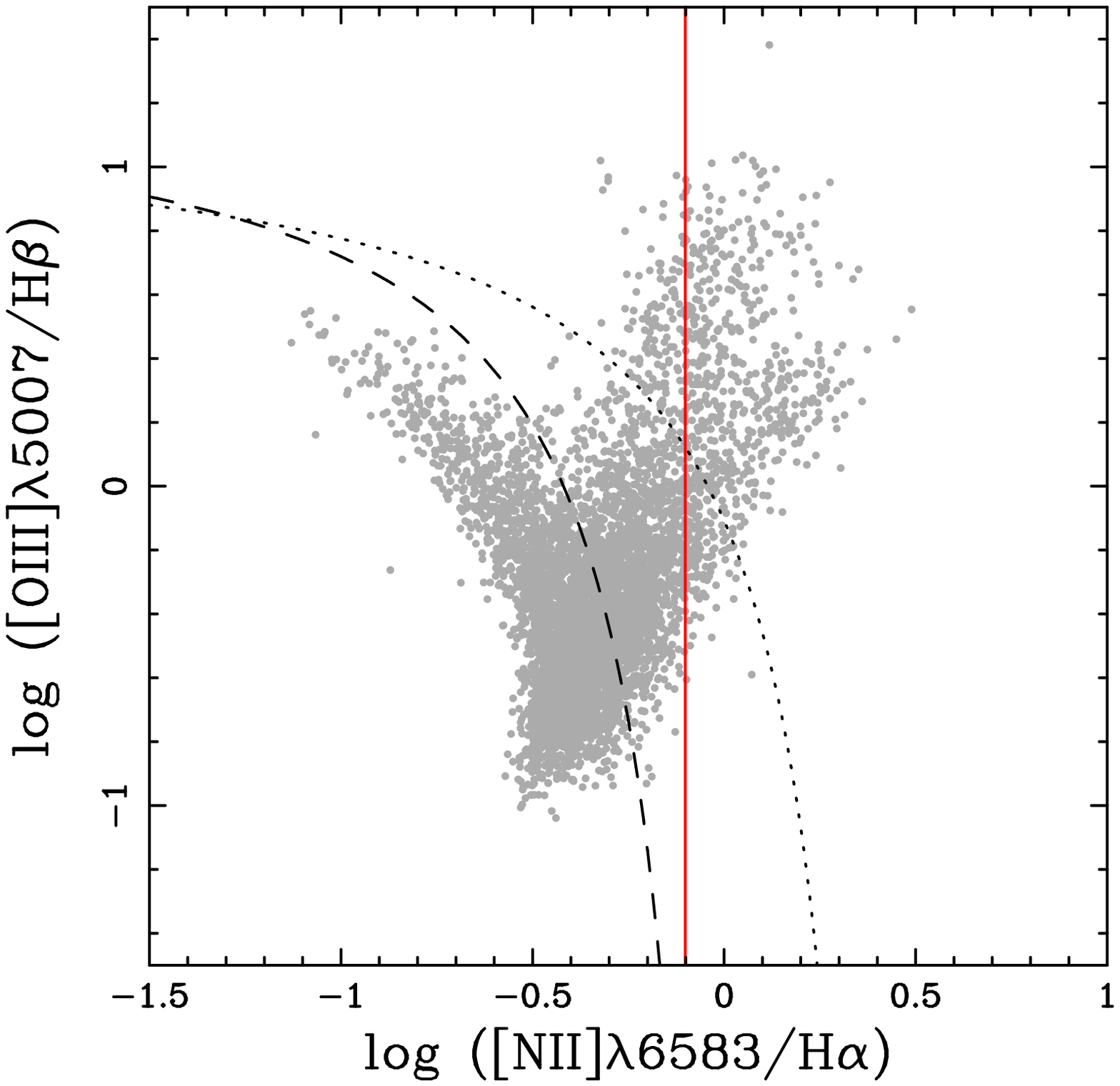}
\caption{
(Left) The optical spectrum of Arp~187 around the H$\alpha$ line. The black/red solid line represents the obtained/fitted spectrum, respectively.
Each gray lines represent the fitted line component, respectively. (Right) The location of Arp~187 (red solid line) in the BPT diagram.
The dotted/dashed line represents the relation of \cite{kau03}/\cite{kew06}, respectively.
Gray dots represent the data points of SDSS DR7 galaxies \citep{aba09}.
}
\end{center}
\end{figure*}

Arp 187 is a merger remnant at the distance of $z=0.04$ 
(and the equivalent luminosity distance of $d_L = 171$~Mpc).  
Using ALMA Cycle~0 observations, \citet{ued14} find two 3~mm continuum components 
located at both sides of the nucleus of Arp~187, spanning 4~kpc.  
These components are identified as small scale radio jets 
from high-resolution VLA archival images at 4.9~GHz and 8.5~GHz, 
which clearly show the morphology of the radio lobes.  
Assuming the jet angle to the line of sight of 90\degree~and 
a typical expansion speed of radio lobes \citep[0.1$c$;][]{mur99, gio09}, 
the kinematic age of the radio-jets is estimated to be $6\times10^4$~yr, which is quite young.  
Although the similar candidates have been found recently 
in the relatively distant universe at $ z > 0.1$ \citep[e.g.,][]{kom06},
as of our knowledge there are no candidates in the local universe with $z<0.1$ 
due to their rare populations. 
Considering the properties above, 
Arp 187 is a good nearby candidate of AGN feedback study where AGN feedback has just started 
but not completed the star formation quenching \citep[e.g., ][ for the molecular gas and the jet interaction study]{mat15}.

While Arp 187 has radio-jets,
it is still not clear whether its AGN activity is still on-going 
due to the absence of previous studies.
The most secure way to identify AGN is a hard X-ray observation, 
because there is no strong bias against absorption up to $\log N_{\rm H} \sim 24.5$
based on theoretical \citep[e.g.,][]{bri11} and observational studies \citep[e.g.,][]{ich12}.  
However, the 70-month integration of \textit{Swift}/BAT 14--195~keV all-sky survey \citep{bau13} 
did not detect Arp 187, constraining the luminosity of $\log L_{14-195 {\rm keV}} \le 43.7$.
This suggests that AGN activity in Arp~187 might be already weaken and/or obscured.
Infrared (IR) observations give us another secure way to find AGN
since the AGN dust torus emission is dominated in mid-IR \citep{ima11, ich12, ich14, ich15}.
Utilizing the \textit{Spitzer} mid-IR imaging data, \cite{ohy15} found AGN signs in the 
southern nucleus of NGC~3256 hosting molecule outflows \citep{sak14}.
\cite{alo12} observed \textit{Spitzer}/IRS 5--38~$\mu$m spectra of LIRGs and 
decomposed the spectra into starburst (SB) and AGN components 
using a SB galaxy template and clumpy torus models \citep{nen08a} 
supported by observations \citep{gan09, ich12, shi13, mar14, asm15}.
In this letter, we report the investigation of AGN activity embedded in Arp~187 
through the optical, IR, and radio-band studies.
Throughout the paper, we adopt $H_{0}=70.0$~km~s$^{-1}$~Mpc$^{-1}$, $\Omega_{M}=0.3$, $\Omega_{\Lambda}=0.7$.

\section{Optical Spectra}
The optical spectra enable us to disentangle type-2 AGN and  starburst galaxy  using the optical line ratios.
Considering AGN have harder spectra than the galaxies, the line ratio with a different ionization potential gives
a good separation between the narrow line region (NLR) gas ionized by AGN and \textsc{Hii} region \citep[called BPT diagram; ][]{bal81, vei87}.
We examined an archival optical spectrum obtained by the 6dF galaxy survey \citep{jon09}.
It is noted that the spectra of 6dF survey are not properly flux calibrated, but the flux ratio of adjacent lines is still useful.
The spectrum covers a range from 3900\AA\ to 7500\AA\ with the fiber aperture of 6.7~arcsec (equivalent to $\sim 5.6$~kpc at $z=0.04$).
Although there is a clear detection of crucial line sets for the BPT diagram including H$\beta$, [OIII]$\lambda5007$, H$\alpha$, and [NII]$\lambda6583$,
one should use caution to obtain H$\beta$ line flux as the additional spectral decomposition is necessary to estimate
how much the H$\beta$ absorption line from the host galaxy affects the observed H$\beta$ emission line.
Therefore, we used [N II]$\lambda6584$/H$\alpha$ only because the concern above should be negligible for these two lines \citep[e.g.,][]{lee11}.
The left panel of figure~1 shows the spectrum of Arp~187 around the H$\alpha$ line. 
The derived flux ratio of [N II]$\lambda6584$/H$\alpha$ is 0.79.
The right panel of figure~1 shows where Arp~187 locates in the BPT diagram.
Above/below the dotted/dashed line is the locus of AGN/starburst galaxy, respectively.
The location between the two lines between \cite{kau03} (dotted) and \cite{kew06} (dashed) is called composite area where the galaxy has an AGN, 
but the strong contribution from the host galaxy.
This shows that Arp~187 belongs to a composite galaxy, LINER, or AGN locus.

One caveat for this result is that the stellar absorption at the H$\alpha$ line could affect the position of BPT diagram.
To check this effect, we measured H$\alpha$ absorption strength from the spectrum template of elliptical galaxy in \cite{kin96}, 
resulting the estimated equivalent width is small with $\sim$1 \AA. Considering H$\alpha$ equivalent
 width of Arp 187 is $\sim$20, the effect is only $\sim$5\%, shifting the ratio of [NII]/Ha
from 0.79 to 0.74. This does not change our main result.
Another caveat is that the observed [N II]$\lambda6584$/H$\alpha$ could be shifted by a radio-jet shock excitation.
\cite{all08} showed that fast shock can mimic the position of the source into the AGN locus in the BPT diagram if the galaxy
emission is fully dominated by shocks. However, in reality, shocks rarely dominate a galaxy's global emission \citep[e.g.,][]{kew13}.
Using the large sample of radio-loud and radio-quiet AGN in the SDSS survey, \cite{kau08} showed that 
both AGN locate remarkably similar positions in the BPT diagram, suggesting that the radio-shock contribution could be
negligible for most of the radio-loud sources in the local universe.
Further checks on the radio-jet contribution will be explored after obtaining the flux calibrated high spatial optical spectra
or Integral Field Unit (IFU).

 \section{IR Spectra}
 \subsection{\textit{Spitzer}/IRS spectra}
 
We obtained \textit{Spitzer}/IRS low-resolution 5.0--38.0~$\mu$m spectra from Cornell Atlas of \textit{Spitzer}/IRS Sources
\citep[CASSIS;][]{lab11}. 
As CASSIS identified Arp~187 as an extended source with a spatial extent of 4.5 arcsec,
it used the extraction aperture width which scales with the spatial extent to account for all of the source's flux.

The obtained IRS spectra include the \textsc{[Oiv]}25.89~$\mu$m line with an ionization potential of $E_{\rm p} = 54.9$~eV \citep{rig09}, which is one of the indicators of AGN
and small contamination from the starburst \citep[$\sim5$\% of the line flux in AGN on average; ][]{per10}.
The spectra of Arp~187 shows a marginal detection of the \textsc{[Oiv]}25.89~$\mu$m line at S/N$\sim 3$, suggesting a weak AGN NLR activity.
The line luminosity of \textsc{[Oiv]}25.89~$\mu$m was obtained with $L_{\rm [OIV]}=6.7\times10^{40}$~erg s$^{-1}$.
Using the large sample of nearby AGN that have both of X-ray and \textit{Spitzer}/IRS spectra, \cite{liu14} showed the
clear luminosity relations between $L_{\rm [OIV]}$ and $L_{\rm 2-10 keV}, L_{\rm 14-195 keV}$.
Using the relation of \cite{liu14},  estimated X-ray luminosities are $L_{\rm 2-10 keV} = 2.5\times10^{42}$~erg s$^{-1}$ and  
$L_{\rm 14-195 keV} = 8.9\times10^{42}$~erg s$^{-1}$. 
\cite{lam10} also reported the relationship between $L_{\rm [OIV]}$ and AGN 13.5~$\mu$m MIR continuum luminosity $L^{\rm (AGN)}_{\rm MIR}$, $L_{\rm 14-195 keV}$
suggesting that the expected $L^{\rm (AGN)}_{\rm MIR}= 1.9\times10^{43}$~erg s$^{-1}$ and  $L_{\rm 14-195 keV}= 1.4\times10^{43}$~erg s$^{-1}$, respectively.
 Note that the intrinsic scatter of each luminosity-luminosity relations is large with $\sim$0.3--0.5 dex.

\begin{figure}
\includegraphics[width=7.9cm]{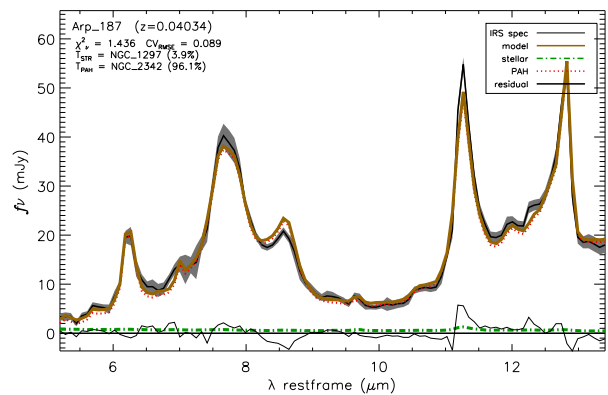}
\caption{
The \textit{Spitzer}/IRS spectra of Arp~187. The brown line represents the model spectra
of STR+PAH component obtained from deblendIRS. The green dashed-dot/red dotted line represents the STR and PAH
component, respectively.
}
\end{figure}

\subsection{Thermal emission from AGN torus}
We decompose the IRS spectra into a combination of stellar direct, host galaxy, and AGN.
We use the IDL routine of \cite{her15} called DeblendIRS.
It prepares a linear combination of three spectral templates of AGN, the stars (``STR''), and the interstellar medium of the host galaxy (``PAH''),
all of which are selected from a large library of IRS spectra from the sources with ``pure-AGN'', ``pure-stellar'', and ``pure-interstellar'' spectra.
A further description of the decomposition method is given in Section 2 of \cite{her15}.
We apply this spectral decomposition routine to the obtained IRS spectra.
We only use the IRS short band (5.0--14.0~$\mu$m) to follow the same manner as \cite{her15}.
They noted that it is difficult to distinguish between the continuum emission of the AGN and
the host at the longer wavelength with $\lambda>14$~$\mu$m.
In addition, the fraction of the total continuum from the AGN decrease drastically at longer wavelength
with the combined effects of a steeply rising emission from dust heated by star formation.

In order to check whether an AGN component contribute to the IR spectra,
we first fit the IRS only with STR+PAH component. The best fitting result is
 achieved with the STR contribution of 3.9\% and PAH contribution of 96.1\%.
 The PAH is the dominant component in the spectra as shown in Figure~2. 
 The resultant reduced $\chi^2$ value is
 $\chi^2/{\rm dof} = 142.3/99 = 1.437$. 
 Then we fit the spectra again with STR+PAH+AGN component.
 The best fit is achieved with the STR contribution of 3.0\%, PAH contribution of 95.5\%,
 and AGN contribution of 1.6\% with the reduced $\chi^2$ value of  $\chi^2/{\rm dof} = 139.8/98 = 1.427$.
 The improvement of the fit based on the F-test shows that the probability is $P=0.196$, which is larger than 0.05, suggesting
that the addition of AGN component does not give significant improvement to explain the spectra.
 Therefore we conclude that the thermal emission from AGN is already weak or diminished in Arp~187.

One caveat is the possibility that the thermal emission from AGN is hidden within the scatter of the IRS spectra.
The averaged S/N at each bin of the spectra with $\sim {\rm S/N}=10$ enables us to give an upper limit of AGN torus thermal emission.
The upper limit of AGN 12~$\mu$m luminosity is $L_{12 \mu {\rm m}}=1.5\times 10^{42}$~erg~s$^{-1}$.
Using the $L_{2-10 {\rm keV}}$--$L_{12 \mu {\rm m}}$ luminosity relations of nearby Seyfert galaxies obtained by \cite{gan09}, 
$L_{2-10 {\rm keV}} = 2.8\times10^{42}$~erg~s$^{-1}$. The estimated BAT X-ray luminosity is $L_{14-195 {\rm keV}} = 5.9\times10^{42}$~erg~s$^{-1}$
using the ratio of 2.1 under the assumption of photon index of $\Gamma=1.9$~\citep{nem11}. This is consistent with the non-detection at \textit{Swift}/BAT survey.
Comparing the luminosities with those expected from \textsc{[OIV]}~25.89~$\mu$m lines are insightful
because the AGN thermal emission originates from the central $<10$~pc, while \textsc{[OIV]}~25.89~$\mu$m line emission would arise from the NLR with $\sim1$~kpc scale.
 The upper limit of $L_{12 \mu {\rm m}}$ is one order of magnitude smaller than the expected $L_{\rm MIR}^{\rm (AGN)}$ from \textsc{[O IV]}25.89~$\mu$m line.
The upper limit of X-ray luminosities are comparable or slightly smaller than the expected value from  \textsc{[O IV]}25.89~$\mu$m lines.

On the intrinsic nuclear AGN activity, final check should be done by direct high sensitive X-ray observations.
\textit{NuSTAR} \citep{har13} and/or \textit{ASTRO-H} \citep{tak14} give us the detailed information including the hard X-ray ($E>10$~keV) band \citep{bri15}.
The spectra can be obtained in 3--70~keV band with 40~ksec exposure up to $N_{\rm H} < 10^{24}$~cm$^{-2}$
under the assumption of photon index of $\Gamma=1.9$ \citep{nem11}.
Even in the Compton-thick case with $N_{\rm H} \le 3\times 10^{24}$~cm$^{-2}$, the detection can be achievable
with the S/N$=4.5$ around 10--20~keV band. Therefore, the future X-ray observations are effective and highly encouraged to
confirm whether or not the AGN activity exists in Arp~187.

Another caveat is the contamination from the jet synchrotron emission to the IRS spectra.
We check the contribution of jet-emission in the IRS band by extrapolating the power-law obtained 
from the radio bands between 80~MHz and 10~GHz from NASA/IPAC Extragalactic Database (NED).
The contribution of the extrapolated IR emission is three orders of magnitude lower than the observed spectra by \textit{Spitzer}/IRS,
 therefore we conclude the jet-contamination in the IRS band for Arp~187 is negligible.

\section{Constraint of AGN activity through the BH fundamental plane}
The fundamental plane of BH activity gives a relation 
among X-ray luminosity, core radio luminosity and the BH mass 
\citep{mer03,hei03,yua14}.
Based on the nucleus $K$ band photometry in \cite{rot04}, Arp 187 has a stellar mass of $M_{\rm stellar}=1.3\times10^{11}$~$M_{\odot}$
with S\'{e}rsic index of $n=4.1$, suggesting that Arp 187 has a bulge dominated galaxy. Therefore, $M_{\rm bulge} \simeq M_{\rm stellar}$.
Using the $M_{\rm BH}$--$M_{\rm bulge}$ scaling relation of \cite{kor13}, the estimated mass $M_{\rm BH}=6.7\times10^8$~$M_{\odot}$ is given.
The central core ($\lesssim 0.42$~arcsec, which is equivalent to $\lesssim 300$~pc) of Arp~187 is not detected in the VLA 4.9~GHz archival image,
and the 3$\sigma$ upper limit of the radio flux is 210~$\mu$Jy, corresponding to $L_{\rm R} \le 3.7\times10^{37}$~erg~s$^{-1}$.
Combining those two parameters and the equation of the updated fundamental plane of \cite{gul09}
enables us to derive the upper limit of 2--10~keV luminosity with 
$L_{2-10 {\rm keV}}<6.6\times10^{36}$~erg~s$^{-1}$.
Some studies suggest that a low-Eddington case might follow the different fundamental relation \citep{yua05}. 
The nucleus of Arp 187 could be in those cases. Applying the relation of \cite{yua05} gives a larger upper limit of
$L_{2-10 {\rm keV}}<4.0\times10^{39}$~erg~s$^{-1}$.
Either case is consistent with the result of very weak or absent of thermal emission of AGN 
discussed in Section~3.

\section{Discussion and Summary}
We discuss the results of our optical, IR, and radio studies.
While Arp~187 hosts small jets with the age of $\sim 6 \times 10^4$ yr,
the non-detection of the radio-core and weak or diminished thermal emission from AGN torus in the IR band prefers the absence 
of an AGN activity in Arp~187, therefore we conclude that the galaxy's central engine has decreased
the energy output by at least a few orders of magnitude within $6\times10^4$ yr.

The results of optical BPT diagram and the marginal detection of \textsc{[O iv]}25.89~$\mu$m indicate
that the NLR of Arp~187 is still alive.
Considering the size of NLR is larger for three to four order of magnitude (1--10~kpc) compared to
the thermal emission region \citep[called torus; $<10$~pc;][]{bur13} of AGN, NLR activity traces the past activity
 at least $10^{4-5}$~yr ago considering the light travel time from the central engine to the NLR \citep[same method was employed by
 the authors of ][]{dad10, kee12}.
This is consistent with the sign of past activity in the radio-jet and its age is $6\times10^4$~yr.
Combining all the results, AGN in Arp~187 was active at least $10^{4-5}$~yr ago, but it is already quenched now and only the extended area leaves the signature of past AGN activity.
The similar type of populations were reported previously.
One is called changing-look AGN. \cite{lam15} reported the fading AGN that experienced over one order of magnitude optical/X-ray flux declining within 10 yr
and drastic disappearance of broad H$\beta$ component. This change might be caused by the declining of accretion rate \citep{eli14}.
Another similar population is called ``Hanny's Voorwerp'', 
which has a faded central engine, but the larger scale NLR emission has been seen as the light echo of AGN
\citep{lin09, kee12, sch15}. These studies of Voorwerps estimate the quenching time scale to $10^{4-5}$ yrs
 via the geometry of their spatial extent. 

One question arises how the sudden quenching of AGN occurs within $10^{4-5}$~yr.
\cite{sch10} discussed the possibility of such short quenching time scale of AGN with an analogy
of X-ray black hole binary study. State changes of GRS 1915+105 with a BH mass of 10~$M_{\odot}$ occur with the time scale of 1~hr.
Extrapolating the BH mass-time variation relation above to Arp~187 with $M_{\rm BH} = 6.7 \times 10^8M_{\odot}$, the state change can occur with the time scale of $\sim10^4$~yr, which is in accordance with the estimated time scale from NLR and radio-jet within an order of magnitude.
This also may support that Arp~187 has experienced the state transition from a high state to a radiatively inefficient state
where the energy is released mainly as kinetic energy \citep{ich77,nar94}.
The presence of a radio-jet also favors the hypothesis above if the launch of the jet is associated with the state change.
Arp~187 is a good nearby candidate for future observations of both central engine with \textit{Nustar, ASTRO-H} and  the host galaxy with ALMA to answer
 how the AGN is going to die and affects the environment of the host galaxy.

\begin{ack}
We are grateful for useful comments from the anonymous referee.
We thank Yoshiki Matsuoka, Tohru Nagao, Ryou Ohsawa, Kouji Ohta, Chris Packham, and Yoshihiro Ueda for valuable comments and discussions.
This research has made use of the NASA/IPAC Extragalactic Database (NED) which is operated by the Jet Propulsion Laboratory, California Institute of Technology, under contract with the National Aeronautics and Space Administration.
This work was partly supported by the Grant-in-Aid  for Scientific Research
40756293 (K.I.).
\end{ack}


\end{document}